\documentstyle[12pt]{article}
\input epsf
\newcommand{\beq}{\begin{eqnarray}}   
\newcommand{\eeq}{\end{eqnarray}}

\newcommand{\gsim}{\lower.7ex\hbox{$
\;\stackrel{\textstyle>}{\sim}\;$}}
\newcommand{\lsim}{\lower.7ex\hbox{$
\;\stackrel{\textstyle<}{\sim}\;$}}

\input epsf
\newcommand{\grpicture}[1]
{
    \begin{center}
        \epsfxsize=200pt
        \epsfysize=0pt
        \vspace{-5mm}
        \parbox{\epsfxsize}{\epsffile{#1.eps}}
        \vspace{5mm}
    \end{center}
}

\begin{document}
\begin{titlepage}
\renewcommand{\thefootnote}{\fnsymbol{footnote}}

\begin{center} \Large
{\bf Institute of Experimental and Theoretical Physics }
\end{center}
\begin{flushright}
ITEP-TH-30/97

\end{flushright}
\vspace*{3cm}

\begin{center}
{\Large \bf  Complex BPS Domain Walls  and Phase Transition in Mass
in Supersymmetric QCD}
\vspace{0.8cm}

{\Large  A. Smilga and A. Veselov} 

\end{center}

\vspace*{2cm}

\begin{abstract}

We study the domain walls connecting different chirally asymmetric
vacua in supersymmetric QCD. We show that BPS -- saturated solutions
exist only in the limited range of mass $m \leq m_* \approx 0.8 
|<{\rm Tr}\ \lambda^2>|^{1/3}$. When $m > m_*$, the domain wall either ceases 
to be BPS -- saturated or disappears altogether. In any case, the properties 
of the system are qualitatively changed.

\end{abstract}

\end{titlepage}

\section{Introduction}

Supersymmetric QCD is the theory involving a gauge vector supermultiplet
$V$ and a couple of chiral matter supermultiplets $Q^f$ belonging to the
fundamental representation of the gauge group $SU(N)$. The lagrangian
of the model reads
\beq
{\cal L} =  \frac{1}{2g^2} \mbox{Tr} \int d^2\theta \ W^2\ + \ 
 \frac{1}{4}\int d^2\theta d^2\bar\theta\ 
\bar S^f  e^V S^f -\left( \frac{m}{4} \int d^2\theta\  
S^{f}S_{f} +\mbox{H.c.}\right)\, ,
\label{LSQCD}
\eeq
where $S_f = \epsilon_{fg} S^g$ (for further conveniences, we have 
changed a sign of mass here compared to the standard convention).
The dynamics of this model is in many respects similar to the dynamics
of the standard (non--supersymmetric) QCD and, on the other hand,
supersymmetry allows here to obtain a lot of exact results \cite{brmog}.

Like in the standard QCD, the axial $U_A(1)$ symmetry corresponding to
the chiral rotation of the gluino field and present in the tree--level
lagrangian (\ref{LSQCD}) is broken by anomaly down to $Z_{2N}$. This
discrete chiral symmetry can be further broken spontaneously down to $Z_2$
so that the chiral condensate $<{\rm Tr}\ \lambda^2>$ is formed. There are
$N$ different vacua with different phases of the condensate
  \beq
  <{\rm Tr}\ \lambda^2> \ =\ \Sigma e^{2\pi i k/N},\ \ \
  \ \ \  k = 0, \ldots, N-1
  \label{cond}
   \eeq
   It was noted recently \cite{Kovner} that on top of $N$ chirally
asymmetric vacua (\ref{cond}), also a chirally symmetric vacuum with
zero value of the condensate exists.

The presence of different degenerate physical vacua in the theory
implies the existence of domain walls --- static field configurations 
depending  only on one spatial coordinate ($z$) which interpolate between
one of the vacua at $ z = -\infty$ and another one at $z = \infty$ and
minimizing the energy functional. As was shown in \cite{Dvali}, in many
cases the energy density of these walls can be found exactly due to the
fact that the walls present the BPS--saturated states. The key
ingredient here is the central extention of the $N=1$ superalgebra 
\cite{Dvali,my}
  \beq
\{ Q^\dagger_{\dot\alpha}Q^\dagger_{\dot\beta}\}
= 4
\left(\vec\sigma\right)_{\dot\alpha\dot\beta}\int
 \, d^3 x \, \vec\nabla  \left\{\left[
-\frac{m}{2} S^{ f} S_{ f} + \frac{1}{16\pi^2}
\mbox{Tr}\ W^2\right]
- \frac{N}{16\pi^2}
\mbox{Tr}\ W^2 \right\}_{\theta = 0} \, ,
\label{cexm}
  \eeq
A domain wall is a configuration where the integral of the full
derivative in the RHS of Eq.(\ref{cexm}) is non--zero so that the standard
$N=1$ SUSY algebra in the wall sector is modified. A BPS--saturated
wall is a configuration preserving 1/2 of the original supersymmetry i.e.
a configuration annihilated by the action of two certain real linear
combinations of the original complex supercharges $Q_\alpha$
\footnote{Such BPS--saturated walls were known earlier in
2--dimensional supersymmetric theories (they are just solitons there)
\cite{Vafa} and were considered also in 4--dimensional theories in stringy
context \cite{ATCQR}. }.

Combining (\ref{cexm}) with the standard SUSY commutator $\{Q_\alpha,
\bar Q^{\dot\beta}\} = 2 (\sigma_\mu)_\alpha^{\dot\beta} P_\mu$ and bearing
in mind that the vacuum expectation value of the expression in the
square brackets in Eq.(\ref{cexm}) is zero due to Konishi anomaly
\cite{Konishi} 
\footnote{See recent \cite{Chib} for detailed pedagogical explanations.},
it is not difficult to show that the energy density of a BPS--saturated
wall in SQCD satisfies a relation
  \beq
   \label{eps}
   \epsilon \ =\ \frac N{8\pi^2} \left|<{\rm Tr}\ \lambda^2>_\infty
    \ -\ <{\rm Tr}\ \lambda^2>_{-\infty} \right|
    \eeq
where the subscript $\pm \infty$ marks the values of the gluino
condensate at spatial infinities.    
    
    The relation (\ref{eps}) is valid {\it assuming} that the wall is
BPS--saturated. However, whether such a BPS--saturated domain wall
exists or not is a non--trivial dynamic question which can be answered
only in a specific study of a particular theory in interest. 

In Ref.\cite{my} this question was studied for the $SU(2)$ gauge group
in the framework of the effective low energy lagrangian due to Taylor,
Veneziano, and Yankielowicz \cite{TVY}. The situation is particularly
simple when the mass parameter $m$ in the lagrangian (\ref{LSQCD}) is
small compared to $\Lambda_{SQCD} \equiv \Lambda $. In this case, chirally
asymmetric vacua are characterized by large values of the matter scalar
field $\chi$. The theory involves two different energy scales, and one
can integrate out heavy fields and to write the Wilsonean
effective lagrangian describing only light degrees of freedom. It is
the lagrangian of the Wess--Zumino model with a single chiral
superfield $X$ and the superpotential
  \beq
{\cal W} = -\frac{2}{3}\frac{\Lambda^5}{X^2} -\frac{m}{2}X^2
\, .
\label{WHiggs}
\eeq
The corresponding potential $U \ =\ |\partial W/\partial \chi |^2$ has
two different non--trivial minima at $<\chi^2> \ =\ \pm \chi_*^2 \ = \ 
\pm \sqrt{4\Lambda^5/3m}$. A domain wall interpolating
between these vacua is BPS--saturated. The solution can be found
analytically \cite{my}
  \beq
\label{wallhgs}
\chi(z) \ =\ \chi_* \frac{1 + i e^{4m(z-z_0)}}{\sqrt{1 + e^{8m(z-z_0)}}}
  \eeq
where $z_0$ is the position of the wall center.

In this approach, we are not able, however, to detect and study a
chirally symmetric vacuum with $<\chi> = 0$ and the corresponding
domain wall. Chirally symmetric vacuum appears when taking into account
also the degrees of freedom associated with the gluon and gluino
fields. The full TVY effective lagrangian is, again, a Wess--Zumino
model involving two chiral superfields $\Phi$ and $X$ with the
superpotential    
\beq
{\cal W} = \frac{2}{3} \Phi^3 \left[ \ln \frac{\Phi^3 X^2}{\Lambda^5} \ -\ 1
\right] - \frac{m}{2} X^2 
\label{WTVY}
\eeq
The corresponding potential for the lowest components $\phi$, $\chi$ of
the superfields $\Phi$, $X$
 \beq
U(\varphi, \chi) \ =\ \left|\frac{\partial W}{\partial \varphi}\right|^2 
+
 \left|\frac{\partial W}{\partial \chi}\right|^2 \ =\ 
4\left| \varphi^2 \ln(\varphi^3 \chi^2) \right|^2 +
\left|\chi\left(m - \frac{4\varphi^3}{3\chi^2} \right)\right|^2
\label{potTVY}
  \eeq
(in the remainder of this and in the next section we will measure
everything in the units of $\Lambda$) has three non--trivial
minima:
  \beq
\phi\ =\ \chi \ =\ 0
\label{min0}
 \eeq
   \beq
\phi \ =\ \left( \frac {3m}4 \right)^{1/6}, \ \ \ \chi \ =\ 
\left(\frac 4{3m} \right)^{1/4}\ ; \nonumber \\     
\phi \ =\ e^{-i\pi/3}\left( \frac {3m}4 \right)^{1/6}, \ \ \ \chi \ =\ 
i\left(\frac 4{3m} \right)^{1/4} 
 \label{minn0}
  \eeq
There are also the minima with inversed sign of $\phi$ and $\chi$, but
they are physically the same as the minima (\ref{minn0}): the vacuum
expectation values of the gauge invariant operators $<{\rm Tr}\ \lambda^2>
\ = \ (32\pi^2/3) <\phi^3>$ and $< s^f s_f> \ =\ <\chi^2>$
($s^f$ is the squark field) are the same.

The effective theory (\ref{WHiggs}) is obtained when the heavy degree
of freedom $\phi$ is frozen in the Born--Oppenheimer spirit so that
$\phi^3 \chi^2  = 1$. (In the opposite limit $m \gg 1$, we can freeze
instead the heavy matter fields $\chi^2 = 4\phi^3/3m$ and arrive at the
Veneziano--Yankielowicz effective lagrangian \cite{VY} for pure
supersymmetric gluodynamics which involves only the field $\phi$.) 
  
 Generally, one should study the theory with the potential
(\ref{potTVY}). The status of this effective theory is somewhat more
uncertain than that of (\ref{WHiggs}) --- for general value of mass,
the TVY effective lagrangian is not Wilsonean; light and heavy degrees
of freedom are not nicely separated. But it possesses all the relevant
symmetries of the original theory 
\footnote{To see that, one should use the amended potential of
Ref.\cite{Kovner} which is ``glued'' of different sectors related
to the different branches of the logarithm; cf. an analogous situation
in the Schwinger model \cite{QCD2}. For our present purposes, these
complications are irrelevant, however.}
 and satisfies the anomalous Ward identities for correlators at zero
momenta. We think that the use of the TVY lagrangian is justified as
far as the vacuum structure of the theory is concerned.

In Ref.\cite{my} the domain walls interpolating between the chirally
symmetric minimum (\ref{min0}) and a chirally asymmetric minimum in
Eq.(\ref{minn0}) were studied along these lines. It was shown that
these walls are BPS--saturated at any value of mass.

In this paper, we study the complex domain walls interpolating between
different minima in (\ref{minn0}). Rather surprisingly, we have found
that the solution of the BPS equation exists only for small enough masses
$m \leq m_* = 4.6705\ldots$ . At larger values of mass, the BPS
equations have no solution. A {\it phase transition} occurs. 
 
 \section{Solving BPS equations.}
 \setcounter{equation}0
BPS equations for the domain wall in a  generalized Wess--Zumino model
with two chiral superfields read \cite{Vafa,Dvali,my}
  \beq
  \partial_z \phi \ =\ \pm \partial \bar W /\partial \bar \phi,
  \ \ \ \ \      \partial_z \chi \ =\ \pm \partial \bar W /\partial
\bar \chi\ 
  \label{BPS}
  \eeq
In our case, the superpotential is given by the expression (\ref{WTVY}).
 Let us choose the positive sign in Eq.(\ref{BPS}) and try to solve it with 
 the boundary conditions
 \beq
 \label{bcwall}
 \phi(-\infty) = \left(\frac{3m}4\right)^{1/6}, \ 
 \phi(\infty)  = e^{-i\pi/3} 
 \left(\frac{3m}4\right)^{1/6}, \nonumber \\  
 \chi(-\infty)  = \left(\frac 4{3m}\right)^{1/4}, \  \chi(\infty)  = i 
 \left(\frac 4{3m}\right)^{1/4} 
 \eeq
 (the negative sign in (\ref{BPS}) would describe the wall going in
  the opposite direction in $z$). The solution of the equations (\ref{BPS})
   with the boundary conditions (\ref{bcwall}) has the fixed energy which
 coincides with (\ref{eps}) after the proper normalization
 \beq
 \label{norm}
 \phi^3 \ =\ \frac 3{32\pi^2} {\rm Tr}\ \lambda^2
 \eeq
 is chosen. 

 Technically, it is convenient to introduce the polar variables 
 $\chi = \rho e^{i\alpha}, \ \ \phi = R e^{i\beta}$.  Then the system 
 (\ref{BPS}) (with the positive sign chosen) can be written in the form
   \beq
   \label{4sys}
   \left\{ \begin{array}{l}
\partial_z \rho \ =\ -m\rho \cos (2\alpha) + \frac{4R^3}{3\rho} \cos (3\beta)
\\
\partial_z \alpha \ =\ m \sin (2\alpha) -  \frac{4R^3}{3\rho^2} \sin (3\beta)
 \\
\partial_z R \ =\ 2R^2 \left[ \cos(3\beta) \ln(R^3\rho^2) -
\sin (3\beta) (3\beta + 2\alpha) \right]  \\
\partial_z \beta \ =\ -2R \left[ \sin(3\beta) \ln(R^3\rho^2) +
\cos (3\beta) (3\beta + 2\alpha) \right] 
\end{array} \right.
   \eeq
The wall solution should be symmetric with respect to its center. Let us seek
for the solution centered at $z=0$ so that 
\beq
\label{sym}
\rho(z)  = \rho(-z), \ R(z) = R(-z),\ \alpha(z)  = \pi/2 - \alpha(-z),
\ \beta(z)  = -\pi/3 - \beta(-z)
\eeq
Indeed, one can be easily convinced that the Ansatz (\ref{sym}) goes through
the equations (\ref{4sys}).

The system (\ref{BPS}) has one integral of motion \cite{Chib}:
\beq
\label{ImW} 
{\rm Im}\ W(\phi, \chi) \ =\ {\rm const}
\eeq
Indeed, we have
$$
\partial_z W = \frac{\partial W}{\partial \phi} \partial_z \phi  \ + \ 
\frac{\partial W}{\partial \chi} \partial_z \chi \ =  \\ 
\left| \frac{\partial W}{\partial \phi} \right |^2 \ +\ 
\left| \frac{\partial W}{\partial \chi} \right|^2 \ =\ \partial_z \bar W
$$
It is convenient to solve the equations (\ref{4sys}) numerically on the
half--interval from $z=0$ to $z = \infty$. The symmetry (\ref{sym}) 
dictates $\alpha(0) = \pi/4,\ \beta(0) = -\pi/6$. The condition (\ref{ImW})
[in our case ${\rm Im}\ W(\phi, \chi) = 0$ due to the boundary conditions
(\ref{bcwall})] implies
  \beq
  \label{Rrho}
\frac{4R^3}3 \left[ \ln(R^3\rho^2) - 1\right] + m\rho^2 \ =\ 0  
   \eeq
Thus, only one parameter at $z=0$ [say, $R(0)$] is left free. We should fit
it so that the solution would approach the complex minimum in Eq.(\ref{minn0})
at $z \to \infty$.

It turns out that the solution of this problem exists, but only in the limited
range of $m$. If $m \geq m_* = 4.6705\ldots$, the solution {\it misses} the 
minimum no matter what the value of $R(0)$ is chosen. This is illustrated in 
Fig. 1 where the ``mismatch parameter''
\beq
\Delta \ =\  \min_{R(0)}  \min_z 
\sqrt{ \left|\chi(z) -  i(4/3m)^{1/4}\right|^2 + 
\left|\phi(z) - e^{-i\pi/3}(3m/4)^{1/6}\right|^2}
\label{mism}
\eeq
 is plotted (in a double logarithmic scale) as a function of mass.
 The dependence $\Delta(m)$ fits nicely the law
  \beq
  \Delta(m) \ =\ 0.56 (m - m_*)^{0.44}
  \eeq
It smells like a critical behavior but, as $\Delta$ is not a physical quantity,
we would not elaborate this point further right now.

\begin{figure}
\grpicture{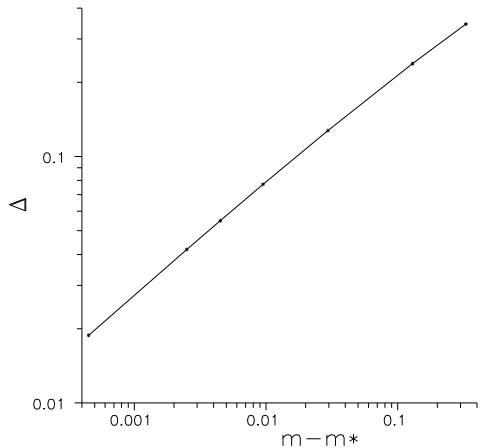}
\caption{ Mismatch parameter $\Delta$ as a function of mass}
\label{mmtch}
\end{figure}

 At $m = 4.6705$ or at smaller values of mass, the solution exists, however.
 The profiles of the functions $\rho(z)$ and $R(z)$ for three
 values of mass: $m = 0.2,\ m = 2.0,$ and $m = 4.6705$ in the whole 
 interval $-\infty <
 z < \infty$  are plotted in Figs. 2,3. For small values of m,
  the solution approaches, as it should, the analytic solution 
 (\ref{wallhgs}) (with $\phi = \chi^{-2/3}$) found in Ref.\cite{my},
so that $\rho(z)$ and $ R(z)$ stay constant.

\begin{figure}
\grpicture{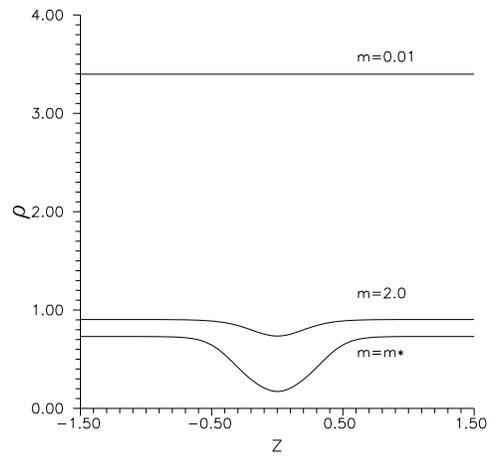}
\caption{ $\rho(z)$ for different  masses.}
\label{rhos}
\end{figure}

\begin{figure}
\grpicture{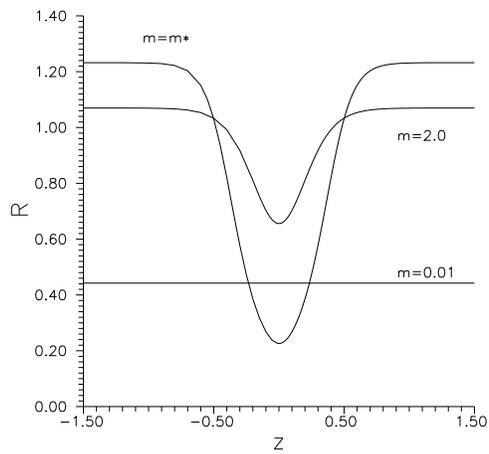}
\caption{$R(z)$ for different masses.}
\label{Rs}
\end{figure}

 In Fig. 4 we plotted the dependence of $R(0)$ on $m$. We see that $R(0)$ has
 a singular behavior at the phase transition point, but stays non-zero.

\begin{figure}
\grpicture{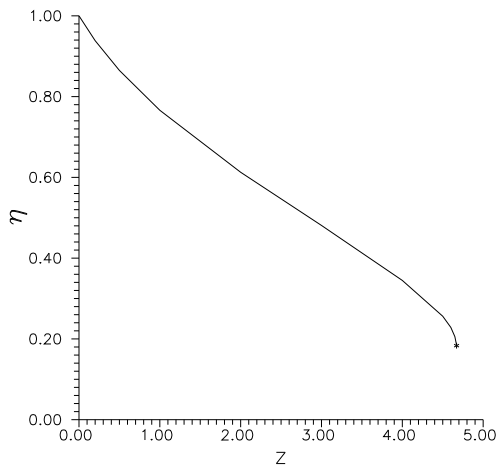}
\caption{The ratio $\eta = R(0)/R(\infty)$  as a function of mass}
\label{R0s}
\end{figure}

 \section{Discussion.}
 \setcounter{equation}0
 We have found that the properties of the system are drastically changed 
 at $m = 4.6705\ldots \Lambda$. It makes sense to express the result in
 invariant terms and to trade $\Lambda$ for an invariant physical quantity
 such as the gluino condensate $\Sigma = | <{\rm Tr}\ \lambda^2>| $ in a
 chirally asymmetric vacuum. From Eqs.(\ref{minn0}), (\ref{norm}), we obtain
  \beq
\Sigma \ =\ \frac{16\pi^2}{\sqrt{3}} m^{1/2} \Lambda^{5/2}
\label{SigLam}
  \eeq
 Thus, the phase transition occurs at
 \beq
 \label{mSig}
m_* \ \approx\ 0.8 \Sigma^{1/3}
 \eeq
The particular numerical value (\ref{mSig}) was obtained by studying the
effective TVY lagrangian with the potential (\ref{potTVY}) and 
the standard kinetic term $|\partial \phi|^2 + |\partial \chi|^2$. We cannot
claim that the phase transition in the theory of interest (\ref{LSQCD})
would occur at exactly the same value of mass. We believe, however, that
the value of $m_*$ in the supersymmetric QCD should be close to (\ref{mSig}).

We hasten to comment that it is not a phase transition of habitual
 thermodynamic variety. In particular, the vacuum energy is zero both
  below and above the phase transition point --- supersymmetry is never
  broken here. Hence $E_{vac} (m) \equiv 0$ is not singular at $m = m_*$.

  Some similarities may be observed with the 2--dimensional Sine--Gordon
  model where the number of the states in the spectrum depends on the 
  coupling constant $\beta$ so that the states appear or disappear at some
  critical values of $\beta$ \cite{SG}. May be a more close analogy can be
  drawn with the $N=2$ supersymmetric Yang--Mills theory. The spectrum of
the system depends on the Higgs expectation value 
$u\  = \ <{\rm Tr} \ \varphi^2>$.
A study of the exact solution of the model due to Seiberg and Witten 
\cite{SW} displays the existence of a ``marginal stability curve'' in
the complex $u$--plane \cite{stab}. When crossing this curve, the spectrum
pattern is qualitatively changed.

To understand better the physics of the phase transition in  supersymmetric
QCD, one has to study in more details what happens in the region $m > m_*$.
There are two logical possibilities:
\begin{itemize}
\item The complex domain wall solution may still exist at larger masses, but 
this solution cannot be BPS--saturated anymore.
\footnote{For detailed discussion of a variety of model examples where the 
walls are or are not BPS--saturated, see \cite{Chib}. Note that our 
observation
 contradicts a {\it conjecture} of Ref.\cite{Chib} that the BPS solutions
 cannot appear or disappear at finite values of parameters in superpotential.}
In this case, it would be very interesting to study what happens in the limit
$m \to \infty$ when the matter fields decouple and the theory is reduced to
 the    pure $N=1$ supersymmetric Yang--Mills theory. Do the domain walls
 interpolating between different chirally asymmetric vacua survive in this
 limit ?
 \footnote{The complex domain walls in the pure $N=1$ SYM theory were 
 discussed (assuming that they are there, of course) in recent \cite{Witten} 
 in the context of D--brane dynamics.}
\item The complex domain wall solution can disappear altogether at $m > m_*$.
 \end{itemize}

 The question of the existence and, if the solutions are there, of the
  properties of the complex domain walls at large masses is now under study.
  But in any case, it is clear now that there {\it is} no smooth transition
  between the weak coupling Higgs regime which is realized at small masses 
  in chirally asymmetric phases and the strong coupling regime at large
  mass values.

\vspace{.5cm}

{\bf Acknowledgments}: \hspace{0.2cm} 
 A.S. acknowledges illuminating discussions with A. Kovner and M. Shifman.
This work was supported in part  by the RFBR--INTAS grants 93--0283, 94--2851,
and 95--0681, by the RFFI grants 96--02--17230 and 97--02--17491, by
the RFBR--DRF grant 96--02--00088,
by the U.S. Civilian Research and Development Foundation under award 
\# RP2--132, and by the Schweizerishcher National 
Fonds grant \# 7SUPJ048716.


\vspace{0.2cm}

\begin{thebibliography} {99}


\bibitem{brmog} 
V. Novikov, M. Shifman, A. Vainshtein, and V. Zakharov,
{\it Nucl. Phys. } {\bf B229} (1983) 407; {\it
Phys. Lett.} {\bf B166} (1986) 334;
I. Affleck, M. Dine and N. Seiberg,
{\it  Nucl. Phys.} {\bf B241} (1984) 493; {\bf B256} (1985) 557.
G. Rossi and G. Veneziano,
{\it Phys. Lett.} {\bf 138B} (1984) 195;
D. Amati, K. Konishi, Y. Meurice, G. Rossi and G. Veneziano,
{\it Phys. Rep.} {\bf 162} (1988) 557;
M. Shifman,
{\it Int. J. Mod. Phys.} {\bf A11} (1996) 5761.


\bibitem{Kovner}
A. Kovner and M. Shifman, hep-th/9702174 [Phys. Rev. D, to appear].


\bibitem{Dvali}
G. Dvali and M. Shifman,  {\it Phys. Lett.}  {\bf B396} (1997) 64;
 hep-th/9611213 [Nucl. Phys. B, to appear].

\bibitem{my} A. Kovner, M. Shifman, and A. Smilga, hep-th/9706089.

\bibitem{Vafa} E. Witten and D. Olive, {\it Phys. Lett.} {\bf B78}
(1978) 97; S. Cecotti and C. Vafa, {\it Commun. Math. Phys.} {\bf 158}
(1993) 569.

\bibitem{ATCQR}
E. Abraham and P. Townsend, {\it Nucl. Phys.} {\bf B351} (1991) 
313; M. Cvet\v{c}, F. Quevedo, and S.-J. Rey, {\it Phys. Rev. Lett.}
{\bf 67} (1991) 1836.

\bibitem{Konishi}
K. Konishi, {\it Phys. Lett.} {\bf B135} (1984) 439;
K. Konishi and K. Shizuya, {\it Nuov. Cim.} {\bf A90} (1985) 111.


\bibitem{Chib}
B. Chibisov and M. Shifman, hep-th/9706141

\bibitem{TVY}
T. Taylor, G. Veneziano and S. Yankielowicz,
{\it Nucl. Phys. } {\bf B218} (1983) 493.


\bibitem{VY}
G. Veneziano and S. Yankielowicz,
{\it Phys. Lett.} {\bf 113B} (1982) 231.


\bibitem{QCD2}
A. Smilga, {\it Phys. Rev.} {\bf D54} (1996) 7757.

\bibitem{SG} R. Dashen, B. Hasslacher, and A. Neveu, {\it Phys. Rev.}
{\bf D11} (1975) 3424.


\bibitem{SW}
N. Seiberg and E. Witten,
{\it  Nucl. Phys.}  {\bf B426} (1994) 19; (E) {\bf B430} (1994) 485.

\bibitem{stab} F. Ferrari and A. Bilal, {\it Nucl. Phys.} {\bf B169}
(1996) 387.

\bibitem{Witten} E. Witten, hep-th/9706109.

\end{thebibliography}
\end{document}